\documentclass[12pt,a4paper]{article}
\usepackage{graphicx}
\usepackage{soul}
\usepackage{xcolor}
\usepackage[square,numbers,comma]{natbib}
\usepackage{setspace}
\usepackage{fullpage}

\title{Wide Binaries, Retardation and the External Field Effect}

\author{Asher Yahalom\\
Ariel University, Ariel 40700, Israel\\
e-mail: asya@ariel.ac.il
}

\begin{document}
\maketitle
\newcommand{\abs}[1]{ |#1|}
\newcommand{\beq} {\begin{equation}}
\newcommand{\enq} {\end{equation}}
\newcommand{\ber} {\begin {eqnarray}}
\newcommand{\enr} {\end {eqnarray}}
\newcommand{\eq} {equation}
\newcommand{\eqs} {equations }
\newcommand{\mn}  {{\mu \nu}}
\newcommand{\ab}  {{\alpha \beta}}
\newcommand{\abp}  {{\alpha \beta}}
\newcommand{\sn}  {{\sigma \nu}}
\newcommand{\rhm}  {{\rho \mu}}
\newcommand{\sr}  {{\sigma \rho}}
\newcommand{\bh}  {{\bar h}}
\newcommand{\br}  {{\bar r}}
\newcommand{\mnras} {Monthly Notices of the Royal Astronomical Society}
\newcommand{\apj} {Astrophysical Journal}
\newcommand{\prd} {Physical Review D}
\newcommand{\pasp} {Publications of the Astrophysical Society of Japan}
\newcommand{\prl} {Physical Review Letters}
\newcommand{\apjs} {Astrophysical Journal Supplement}
\newcommand{\aj} {Astronomical Journal}
\newcommand{\aap} {Astronomy and Astrophysics}
\newcommand {\er}[1] {equation (\ref{#1}) }
\newcommand {\ern}[1] {equation (\ref{#1})}
\newcommand {\Ern}[1] {Equation (\ref{#1})}
\thispagestyle{empty}
\begin{abstract}
Recently a low accleration gravitational anomaly was reported for wide binaries, that is a system of two binary stars which are seperated by more than five killo astronomical units (kau). The increase in gravitation force was reported to be about $1.37$ greater than Newtonian gravity. At the same time binaries which are not wide that is of seperation less than one kau where shown to obey standard Newtonian gravity. A possible expalanation for this was given in the frame work of
MOND correction to gravity which is applicable for low accleration. However, it was noticed that the explanation is only adequate in the frame work of Milgromian AQUAL theory which takes into account the "external field effect" that is the effect of the gravitational field of the rest of the galaxy on the binary system. Recently it was shown that many "anomalous gravity" effects can be explained in the frame work of general relativity and its weak field approximation. It was shown that anomolous galaxy rotation curves, anamolous gravitation lensing and the Tully-Fisher relations can be understood in terms of retarded gravity. Moreover, the anamolous mass of galaxy clusters derived from the virial theorem was also shown to be a retarded gravity effect. On the other hand it was shown that retarded effects are not extremely important in the solar systme and
the needed correction to the anomalous shift of Mercury's perihelion are connected with the motion
of the solar center of mass with respect to the sun and not to retaradation. It is thus desirable to investigate if the wide binary accleration gravitational anomaly can be also explained in the frame-work of Einstein-Newton general relativity. We show that the scale of five kau seperation arises naturally within such a theory if the effect of the galaxy on the wide binary is considered (a Newtonian external field effect) suggesting the anomally found can be explained without the need to invoke theoretical modifications of the accepted theory.
\end{abstract}

\section {Introduction}

It was suggested \cite{Chae2024} that the behaviour of gravity at low accelerations can be directly probed by wide binaries (widely separated stars and thus long-period, but still gravitationally bound binary stars) since any imaginable dark matter in the Milky Way may not deform their internal dynamics (e.g., \cite{hernandez2012,banik2018,pittordis2018,banik2019,pittordis2019,hernandez2019,
elbadry2019,clarke2020,hernandez2022,pittordis2023,hernandez2023}). Thus using the recent data provided by the Gaia satellite multiple statistical analyses have been reported with the purpose of detecting gravity effects which differ from Newtonian predictions (\cite{chae2023a,banik2024,chae2024,hernandez2024}) based on the Gaia data release 3 \cite{dr3}.

Chae \cite{chae2023a} has suggested an algorithm \cite{chae2023b} that calculates the probability distribution of a kinematic acceleration assuming it is of the form $g=v^2/r$ with respect to the Newtonian acceleration $g_N$ between the two stars where $v$ is the relative velocity and $r$ is the separation in the three-dimensional real space, and compares it with a naive Newtonian prediction. One key aspect of this algorithm is that the occurrence rate ($f_{\rm{multi}}$) of multiplicity higher than two (i.e.\ harboring hidden additional components) can be self-calibrated at a high enough acceleration and checked at another high acceleration. Through this algorithm, to be referred to as the "acceleration-plane analysis", \cite{chae2023a} found that the observed acceleration started to get boosted from the Newtonian prediction for $g_N \le 10^{-9}$~m~s$^{-2}$ with a boost factor of $\approx 1.4$ for $g_N \le 10^{-10}$~m~s$^{-2}$, at an extremely high ($>5\sigma$) statistical significance. Chae\cite{chae2023a} considered various samples by varying selection criteria and noted that the low-acceleration gravitational anomaly persisted.

Chae \cite{chae2024} further considered a sample of statistically pure binaries (i.e.\ the limiting case of $f_{\rm{multi}}=0$) for an independent test. For this sample Chae \cite{chae2024} employed another algorithm to be referred to as "stacked velocity profile analysis" in addition to the acceleration-plane analysis. The stacked velocity profile analysis compares the observed distribution of the sky-projected relative velocities against the sky-projected separations with the corresponding Newtonian prediction. Chae \cite{chae2024} found that the results for the pure binary sample through the two independent algorithms agreed well with each other and the \cite{chae2023a} results for general or "impure" samples. It follows that the gravitational modifications at a small acceleration does not depend even on a significant variation of the sample selection criteria, that is between $f_{\rm{multi}}\sim 0.5$ \citep{chae2023a} and $f_{\rm{multi}}= 0$ \citep{chae2024}.

Hernandez \cite{hernandez2024} also performed a statistical analysis of the distribution of velocities in a pure binary sample that was selected by the same author \cite{hernandez2023}. Hernandez defined the normalized velocity on the sky plane \cite{pittordis2018,banik2018} by:
\begin{equation}
  \tilde{v} \equiv \frac{v_p}{v_c(s)},
  \label{eq:vtilde}
\end{equation}
$v_p$ is the observed sky-projected relative velocity between the pair and $v_c(s)$ is the theoretical Newtonian circular velocity at the sky-projected separation $s$. The Hernandez \cite{hernandez2024} algorithm and sample are completely independent of Chae's \cite{chae2023a,chae2024} although the data is from the same Gaia DR3 database. Hernandez \cite{hernandez2024} obtained a gravitational anomaly that was well consistent with those obtained by Chae \cite{chae2023a,chae2024}.

The papers of Chae and Hernandez \cite{chae2023a,chae2024,hernandez2024} show that the gravitational acceleration is increased by $\approx 1.3 - 1.5$ and the relative velocity is boosted by a factor of $\approx 1.2$ for $s\ge 5$~kau. This was interpreted as connected to the MONDian framework since in this case $g_{\rm{N}}\le 10^{-10}$~m~s$^{-2}$ and the MONDian acceleration figure of merit is:
$a_0 \simeq 1.2~10^{-10}$~m~s$^{-2}$. Despite these repeating results from different samples of wide binary star systems with different methods, Banik \cite{banik2024} claimed an opposite conclusion based on another statistical method and doubted the data quality control in Chae's \cite{chae2023a} samples.

However, Hernandez \cite{hernandez2024a} criticised the claims of Banik. The main fault according
to Hernandez is the exclusion by Banik of the Newtonian-regime ($g_N > 10^{-9}$~m~s$^{-2}$ or $s< 2$~kau) binaries that are essential for an accurate determination of $f_{\rm{multi}}$. Then, Banik employed a  distribution  fitting  (number count) of $\tilde{v}$ (Equation~\ref{eq:vtilde}) in cells in an attempt to simultaneously constrain gravity, $f_{\rm{multi}}$, and additional parameters but without the Newtonian-regime data. In such a method they improperly defined cells smaller than the errors of $\tilde{v}$ although Banik relies on number counts in cells \cite{hernandez2024a}.

The concern of data quality \cite{banik2024} due to uncertainty of $\tilde{v}$ has no consequence for the analysis of Chae \cite{chae2023a} because $\tilde{v}$ is not used, but probability distributions of $g$ and $g_N$ are directly derived on the acceleration plane with sufficiently precise projected relative velocities, and considering ranges of parameters such as eccentricity, inclination, and orbital phase in deprojection to the three-dimensional space. The pure binary sample of Chae \cite{chae2024} admits high signal-to-noise ratios ($S/N\ge 10$) for $\tilde{v}$. The acceleration-plane analysis returns similar results for the gravitational anomaly for $g_N \le 10^{-9}$~m~s$^{-2}$.

From an empirical perspective general relativity (GR) is known to be verified by many different types of observations \cite{Einstein2,Edd,Weinberg,MTW,Narlikar}.
However, currently Einstein's general relativity  is at a rather difficult position. It has much support from observational evidence while also having serious challenges. The observational verifications it has gained in both cosmology and astrophysics are in doubt due to the fact that it needs to include unconfirmed ingredients, dark matter and energy, in order to achieve successes on the larger scales of galaxies, clusters of galaxies and universe as a whole. In most cases the
unconfirmed ingredient is used while at the same time practitioners neglect a major ingredient of general relativity, the phenomenon of retardation, that negates Newtonian action at a distance.

Indeed, the dark matter enigma has not only been with the astronomical community since the 1930s (or perhaps even since the 1920s when it was known as the question of missing mass), but it has become prevalent as more  dark matter (and a more serious neglect of retardation) has had to be postulated on larger scales as those scales were scrutinized. A very detailed and costly forty-years underground and accelerator search failed to prove its existence. The dark matter enigma has become even more problematic in recent years as the Large Hadron Collider failed to find any super symmetric particles, not only the astro-particle community's preferred form of dark matter, but an essential ingredient that string theory needs, and it is that same string theory that is expected to quantize gravity.

As early as 1933, Zwicky noted that a group of galaxies within the Comma Cluster have velocities that are significantly higher than that predicted by the virial calculations based on Newtonian theory  \cite{zwicky,DarkMatterMatter}.  He calculated that the amount of matter required to account for the velocities could be 400 times greater than that of visible matter (this was later mitigated somewhat). Of course, if Zwicky would have use the retarded gravity version of the virial theory \cite{YaRe10} no significant problem would arise.  In 1959, on a smaller galactic scale, Volders observed that stars in the outer rims of a nearby spiral galaxy (M33) do not move as they should \cite{volders}. That is to say that the velocities do not decrease as  $1/\sqrt{r}$.
This discrepancy was further established in later years. During the seventies Rubin and Ford \cite{rubin1,rubin2} have shown that for a rather large sample of spiral galaxies the velocities at the outer rim of galaxies do not decrease. Rather, in the general case, they attain a plateau (or continue increasing) at some velocity different for each galaxy. In previous works it was shown that such velocity curves can be deduced directly from GR if retardation is not neglected. The~derivation of the retardation effect is described in previous publications ~\cite{YaRe1,ge,YaRe2,Wagman,Wagman2,YaRe3,YaRe4,YaRe5}. The mechanism is strongly connected to the dynamics of the density of matter inside the galaxy, or more specifically to the densities' second derivative. The density can change due to the depletion of gas in the galaxies surrounding (in
which case the second derivative of the galaxies' total mass is negative \cite{YaRe3}) but can also be affected by dynamical processes involving star formation and supernovae explosions \cite{Wagman,Wagman2}. It was determined that all possible processes can be captured by three different length scales: the typical length of the density gradient, the typical length of the velocity field gradient and the dynamical length scale. It is the shortest among those length scales that determine the significance of retardation \cite{lensing}.

The famous relation of Tully and Fisher \cite{TF} connecting the baryonic mass of a galaxy to the fourth power of its rim velocity can also be deduced from retarded gravity \cite{YaRe6}.

Retarded gravity does not affect only slowly moving particles but also photons.
Although the mathematical analysis is slightly different in both cases \cite{lensing,lensing2}, it is concluded that the apparent "dark mass" must be the same as in the galactic rotation curves.

While the standard dark matter paradigm may still prevail, the current situation is alarming enough to contemplate the possibility that this prevailing paradigm might at least be reconsidered.
Thus there is indeed room for the present suggestion. Unlike other theories that suggest modifying general relativity such as Milgrom's MOND \cite{Milg1983}, Mannheim's Conformal Gravity \cite{Mann93,ManExt} or Moffat's MOG \cite{Moffat}, the current approach does not do so. We adhere strictly to Occam's razor, as suggested by Newton and Einstein. It seeks to replace dark matter
with effects within the standard General Relativity itself. Notice, however, that the connection between retardation and MOND was recently elucidated \cite{YaRe11}, showing in what sense low acceleration MOND criteria can be derived from retardation theory and how MOND interpolation function can be a good approximation to retarded gravity.

We emphasize that appreciable retardation effects do not require that velocities of matter in the galaxy are high (although this may help), in fact the vast majority of galactic bodies (stars, gas) are slow with respect to the speed of light. In other words, the ratio of $\frac{v}{c} \ll 1$. Typical velocities in galaxies are ~100 km/s, which makes this ratio $0.001$ or smaller. To obtain
appreciable retardation effects what is needed is a small
typical gradient scale with respect to the size of the system \cite{lensing}.
It was shown that retardation effects may become significant even at low speeds, provided that the distance over a typical length scale is large enough.

Within the solar system retardation effects are not appreciable \cite{YaRe9,YaRe9b}.
However, galaxies' velocity curves indicate that the retardation effects cannot be neglected beyond a certain distance \cite{Wagman,Wagman2,YaRe3}.
The purpose of this study is to establish the situation in interim gravitational system that is wide binaries, in which distances are larger than the typical distances in the solar system but smaller than the typical size of a galaxy.

\section {Beyond the Newtonian Approximation}

Retarded gravity  can be obtained from the weak field approximation to general relativity \cite{YaRe3}. The metric perturbation $h_{00}$ can be given in term of a retarded potential $\phi$
as follows \cite{YaRe3,lensing}:
\beq
\phi =  -G \int \frac{\rho (\vec x', t-\frac{R}{c})}{R} d^3 x',
   \quad  \phi \equiv \frac{c^2}{2} h_{00},  \quad h_{00} = \frac{2}{c^2} \phi
 \label{photonequastartpoint}
\enq
In the above $G$ is the gravitational universal constant, $\vec x$ is where the potential is measured, $\vec x'$ is the location of the mass element generating the potential,
$\vec R \equiv \vec x - \vec x', \ R \equiv |\vec R|$, and $\rho$ is the mass density.
The duration $\frac{R}{c}$ for galaxies may be a few tens of thousands of years, but~can be considered short in comparison to the time taken for the galactic density to change significantly.
Similarly for clusters of galaxies the duration $\frac{R}{c}$ for galaxies may be a few tens of millions of years, but~can be considered short in comparison to the time taken for the galactic cluster density to change significantly.  Thus, we can write a Taylor series for the density:
\beq
\rho (\vec x', t-\frac{R}{c})=\sum_{n=0}^{\infty} \frac{1}{n!} \rho^{(n)} (\vec x', t) (-\frac{R}{c})^n,
\qquad \rho^{(n)}\equiv \frac{\partial^n \rho}{\partial t^n}.
\label{rhotay}
\enq
By inserting Equations~(\ref{rhotay}) into Equation~(\ref{photonequastartpoint}) and keeping the first three terms, we will obtain:
\beq
\phi = -G \int \frac{\rho (\vec x', t)}{R} d^3 x' +  \frac{G}{c}\int \rho^{(1)} (\vec x', t) d^3 x'
- \frac{G}{2 c^2}\int R \rho^{(2)} (\vec x', t) d^3 x'
\label{phir}
\enq
The Newtonian potential is the first term:
\beq
\phi_N = -G \int \frac{\rho (\vec x', t)}{R} d^3 x'
\label{phiN}
\enq
the~second term does not contribute to the force affecting subluminal particles as its gradient is null and~the third term is the lower order correction to the Newtonian potential:
\beq
\phi_r = - \frac{G}{2 c^2} \int  R \rho^{(2)} (\vec x', t) d^3 x'
\label{phir2}
\enq
The geodesic equation for a any "slow" test particle moving under the above space-time metric can be approximated \cite{YaRe3} using  the force per unit mass as follows:
\beq
 \frac{d \vec v }{d t}  = \vec g, \qquad \vec g \equiv  - \vec \nabla \phi
 \label{equmotion}
\enq
The total force per unit mass is thus:
\ber
\vec g &=& \vec g_{N} + \vec g_{r}
\nonumber \\
 \vec g_{N} &\equiv& -  \vec \nabla \phi_N =
  -  G  \int \frac{\rho (\vec x',t)}{R^2} \hat R d^3 x',  \quad \hat R \equiv \frac{\vec R}{R},
\nonumber \\
 \vec g_{r} &\equiv& - \vec \nabla \phi_r =  -\frac{G}{2c^2} \int  \rho^{(2)} (\vec x', t) \hat R d^3 x'
\label{Fr}
\enr
Now consider a point particle which has a mass $m_j$ and is located at $\vec r_j (t)$, such
a particle will have a mass density of:
\beq
\rho_j = m_j \delta^{(3)} (\vec x' - \vec r_j (t))
\label{rhoj}
\enq
in which $\delta^{(3)}$ is a three dimensional Dirac delta function. This particle will
 cause a Newtonian potential:
\beq
\phi_{Nj} = -G \frac{m_j}{R_j (t)}, \qquad \vec R_j (t) = \vec x - \vec r_j (t),
\qquad  R_j (t) = |\vec R_j (t)|
\label{phiNj}
\enq
and a retardation potential of the form:
\ber
\phi_{rj} &=& - \frac{G m_j}{2 c^2} \frac{\partial^2 }{\partial t^2} R_j (t)
= \frac{G m_j}{2 c^2} \left( \hat R_j \cdot \vec a_j - \frac{\vec v_j^2 - (\vec v_j \cdot \hat R_j)^2}{R_j (t)} \right),
\nonumber \\
~ \hat R_j &\equiv& \frac{\vec R_j}{R_j},
~ \vec v_j \equiv \frac{d \vec r_j}{dt},
 ~ \vec a_j \equiv \frac{d \vec v_j}{dt}.
\label{phirj}
\enr
Thus any point particle moving at the vicinity of particle $j$ will be accelerated as follows:
\ber
\vec g_j &=& \vec g_{Nj} + \vec g_{rj}
\nonumber \\
 \vec g_{Nj} &=& -  \vec \nabla \phi_{Nj} = -G \frac{m_j}{R_j^2} \hat R_j,~
 \vec g_{rj} = - \vec \nabla \phi_r =  \frac{G m_j}{2 R_j^2 c^2}
  \left(R_j \vec a_{\perp j}+ \hat R_j \vec v_{\perp j}^2 - 2 (\vec v_{j} \cdot \hat R_j) \vec v_{\perp j}\right)
  \nonumber \\
 \vec a_{\perp j} &\equiv& \vec a_j - (\vec a_j \cdot \hat R_j) \hat R_j, \qquad
 \vec v_{\perp j} \equiv \vec v_j - (\vec v_j \cdot \hat R_j) \hat R_j.
\label{frj}
\enr
Notice the different notation used for the test particle acceleration $\vec g_j$  and the gravity source acceleration $\vec a_j$.

Now consider a point particle of mass $m_k$ which is located at $\vec r_k (t)$, this particle will
feel the force:
\ber
\vec F_{j,k} &=& \vec F_{Nj,k} + \vec F_{rj,k}
\nonumber \\
 \vec F_{Nj,k} &=& -G \frac{m_j m_k }{R_{k,j}^2} \hat R_{k,j},\qquad
 \vec R_{k,j}  \equiv \vec r_k - \vec r_j,
~  R_{k,j}  \equiv |\vec R_{k,j} (t)|,
~ \hat R_{k,j} \equiv \frac{\vec R_{k,j}}{R_{k,j}},
\nonumber \\
 \vec F_{rj,k} &=&   \frac{G m_j m_k }{2 R_{k,j}^2 c^2}
  \left(R_{k,j} \vec a_{\perp j,k}+ \hat R_{k,j}\vec v_{\perp j,k}^2 - 2 (\vec v_j \cdot \hat R_{k,j}) \vec v_{\perp j,k}\right)
  \nonumber \\
 \vec a_{\perp j,k} &\equiv& \vec a_j - (\vec a_j \cdot \hat R_{k,j}) \hat R_{k,j}, \qquad
 \vec v_{\perp j,k} \equiv \vec v_j - (\vec v_j \cdot \hat R_{k,j}) \hat R_{k,j}.
\label{Frjk}
\enr
We notice once again (see \cite{YaRe4}) that while Newtonian forces are prominent at "short" distances the retardation forces are the most significant at large distances in which it drops as $\frac{1}{R}$ and this fact is not related to the Taylor series approximation that we have used here. Now let us consider the gravitational effect of particle $k$ on particle $j$, this is easily calculated by exchanging the indices $j$ and $k$ in the above expression.
As $R_{j,k} = R_{k,j}$ but $\hat R_{j,k} = - \hat R_{k,j}$ it follows
that the Newtonian force satisfies Newton's third law: $\vec F_{Nk,j} = - \vec F_{Nj,k} $, however, since there is no simple relation between the velocity and acceleration of the particle $j$ and $k$
it follows that generally speaking  $\vec F_{rk,j} \neq - \vec F_{rj,k} $ and thus both the retardation  force and the total gravitational force do not satisfy Newton's third law. This is well
known in electromagnetism and discussed in a series of papers \cite{Tuval,YahalomT,Yahalom3,Yahalom4,Yahalom5,Yahalom6}.

\section{Retardation Corrections}

Let us establish the conditions under which the retardation correction is important. We shall write the retardation force in \ern{Frjk} as:
\beq
\vec F_{rj,k} = \frac{1}{2} F_{Nj,k}
  \left(\frac{R_{k,j} \vec a_{\perp j,k}}{c^2}+ \hat R_{k,j}\frac{v_{\perp j,k}^2}{c^2} - 2
  \frac{(\vec v_j \cdot \hat R_{k,j}) \vec v_{\perp j,k}}{c^2}\right), \qquad
  F_{Nj,k} \equiv |\vec F_{Nj,k}|
\label{Fr2}
\enq
If the velocity of particle $j$ is non relativistic that is much smaller than the speed of light we have:
\beq
\vec F_{rj,k} \simeq \frac{1}{2} F_{Nj,k}
  \left(\frac{R_{k,j} \vec a_{\perp j,k}}{c^2}\right).
\label{Fr3}
\enq
It thus follows that this is the correct expression for most gravitational systems including the wide binary stars discussed by Chae \cite{Chae2024}. Thus this retarded force can be neglected
if:
\beq
\frac{1}{2}  \left(\frac{R_{k,j} a_{\perp j,k}}{c^2}\right) \ll 1
\quad \Rightarrow \quad R_{k,j} \ll R_c \equiv \frac{2 c^2}{a_j} < \frac{2 c^2}{a_{\perp j,k}}
\label{Fr4}
\enq
The critical distance $R_c$ is acceleration dependent it will be relatively small for highly accelerated bodies and much larger for non accelerating bodies. Of course to know what the acceleration is one must well define the inertial system otherwise this notion is arbitrary.
The same statement is of course also correct for MOND type theories which also rely on acceleration to calculate the force, and even more so as in the MONDian case a slight acceleration can make a huge difference in force calculations. The other extreme:
\beq
R_{k,j} \geq R_c \qquad {\rm Or} \qquad R_{k,j} \simeq R_c
\label{Fr5}
\enq
signifies the regime in which retardation forces are important. Thus the gravitational interaction between a galactic central star moving at high acceleration and a distant gas atom at the outskirts
of a galaxy is more likely to be affected by retardation than the gravitational interaction of a low acceleration binary star system even if widely separated. We will put this assertion in quantitative terms in later sections.

\section{Binary Systems}

Let us consider a system of $N$ particles, each particle will feel a force generated by all other particles. $\vec F_k$ is the total force acting on particle $k$ by all the other particles:
\beq
\vec F_k \equiv \sum_{j=1, j\neq k}^{N} \vec F_{j,k}
\label{FK2}
\enq
Without the loss of generality we shall separate a binary subsystem from the system of $N$ particles, made out of the particle $1$ and $2$. We thus write the forces acting on particle $1$ and $2$ as:
\ber
\vec F_1 &=& \vec F_{2,1} + \vec F_{e1},
\qquad  \vec F_{e1} \equiv  \sum_{j=3}^{N} \vec F_{j,1}
\nonumber \\
\vec F_2 &=& \vec F_{1,2} + \vec F_{e2},
\qquad  \vec F_{e2} \equiv  \sum_{j=3}^{N} \vec F_{j,2}
\label{FK3}
\enr
If the particles are non-relativistic we may write the following equations of motion:
 \ber
m_1 \vec a_1 &=& \vec F_1 = \vec F_{2,1} + \vec F_{e1},
\nonumber \\
m_2 \vec a_2 &=& \vec F_2 = \vec F_{1,2} + \vec F_{e2}.
\label{eqmotion}
\enr
We shall now define the center of mass and the relative position of the binary particles (stars)
in the customary way:
\beq
\vec R_{cm} \equiv \frac{m_1 \vec r_1 + m_2 \vec r_2}{M}, \qquad
\vec r \equiv \vec R_{2,1} = \vec r_2 - \vec r_1, \qquad M \equiv  m_1+m_2.
\label{Rcm}
\enq
Thus:
\ber
 \vec r_1 &=& \vec R_{cm} +(\frac{m_1}{M} - 1) \vec r  = \vec R_{cm} -\frac{m_2}{M} \vec r ,
\nonumber \\
\vec r_2 &=& \vec R_{cm} +(1 - \frac{m_2}{M}) \vec r  = \vec R_{cm} +\frac{m_1}{M} \vec r.
\label{r12}
\enr
The acceleration of each star in the binary system can be partitioned to a contribution from the
center of mass acceleration and a relative acceleration as follows:
\ber
 \vec a_1  &=&  \ddot{\vec r_1} =  \ddot{\vec R}_{cm} -\frac{m_2}{M} \ddot{\vec r}
 = \vec a_{cm} -\frac{m_2}{M} \vec a, \qquad
 \vec a_{cm} \equiv \ddot{\vec R}_{cm}, \quad \vec a \equiv \ddot{\vec r}
\nonumber \\
\vec a_2  &=&  \ddot{\vec r_2} =  \ddot{\vec R}_{cm} +\frac{m_1}{M} \ddot{\vec r}
 = \vec a_{cm} +\frac{m_1}{M} \vec a.
\label{a12}
\enr
On the other hand we can calculate the center of mass acceleration as:
\beq
\vec a_{cm} = \frac{m_1 \vec a_1 + m_2 \vec a_2}{M} =
\frac{1}{M}[\vec F_{1,2} + \vec F_{2,1} + \vec F_{e1} + \vec F_{e2}]
= \frac{1}{M}[\vec F_{r1,2} + \vec F_{r2,1} + \vec F_{e1} + \vec F_{e2}].
\label{acm}
\enq
In which the last equation sign follows from Newton's third law:
$\vec F_{N1,2} = -\vec F_{N1,2}$.
We can also calculate the relative acceleration as follows:
\beq
\vec a = \vec a_2 - \vec a_1 =
\vec g_{1,2} - \vec g_{2,1} + \vec g_{e2} - \vec g_{e1}, \qquad
\vec g_{j,k} \equiv \frac{\vec F_{j,k}}{m_k}, \quad \vec g_{ek} \equiv \frac{\vec F_{ek}}{m_k}.
\label{a}
\enq

\section{Orders of Magnitude}

\begin{figure}
 \centering
\includegraphics[width=0.5\columnwidth]{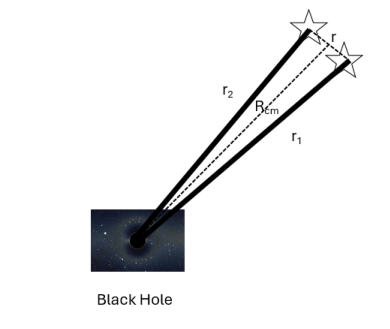}
 \caption{A schematic binary system. The distance $r$ is much exaggerated with respect to $R_{cm}$. }
 \label{schembin}
\end{figure}

We shall assume that the center of the galaxy is the same with the origin of axis of the inertial system. As in Chae \cite{chae2023a,chae2024}, we consider a sample of binaries within $200$ pc from the Sun included in the El-Badry \cite{elbadry2021} database of $1.8$ million binary candidates that are free of clusters, background pairs, and resolved ($1''$) triples or higher-order multiples. The sample of $81,880$ binary candidates defined by Chae \cite{chae2023a} has the range of relative distances of $r_{min}= 0.05$~kau  $ < r=|\vec r| < r_{max}=50$~kau.
Now $R_{cm} = |\vec R_{cm}|$ for each of the binary systems is the same order of magnitude as the distance of the sun from the galactic center which is about $R_{sun} \simeq 26,660$~ly $\simeq 8,174$~pc $\simeq 2.5~10^{20}$~m. The relative distance of the stars in the binary system is much smaller as $r_{max} = 50$~kau $\simeq 0.24$ pc and the relative distance $r$ is smaller than this number even for wide binaries. Thus for each binary system we have:
\beq
r \ll R_{cm},r_1,r_2, \qquad \epsilon \equiv \frac{r}{R_{cm}} < \frac{r_{max}}{R_{cm}}
\simeq 3~10^{-5} \ll 1.
\label{orofmag}
\enq
This leads to the picture depicted in figure \ref{schembin}.
Now the sun circulates the center of the galaxy with a velocity of: $v_{sun} \simeq 251 ~ 10^3$ m/s
leading to a center of mass acceleration of about:
\beq
a_{cm} \simeq a_{sun} \equiv \frac{v_{sun}^2}{R_{sun}} \simeq 2.5~10^{-10}~{\rm m/s^2}.
\label{asun}
\enq
A number not far from the MONDian acceleration scale of $a_0 \simeq 1.2~ 10^{-10} ~{\rm m/s^2}$. The relative acceleration will of course depend on the distance $r$ between the stars in the binary system, thus for order of magnitude estimation only:  
\beq
a_{min} < a < a_{max}, \qquad 
a_{max} \equiv \frac{G M_\odot}{r_{min}^2} \simeq 2.4~10^{-6}~ {\rm m/s^2},
\quad a_{min} \equiv \frac{G M_\odot}{r_{max}^2} \simeq 2.4~10^{-12}~ {\rm m/s^2}
\label{aordest}
\enq
in which we took as typical mass the solar mass $M_\odot \simeq 2~ 10^{30}$ kg.
It follows that according to \ern{a12} there is a difference between close binaries in which
the relative acceleration of the system $a$ dominates over $a_{cm}$ and wide binaries in which the opposite may be true. Curiously enough the border line is at about the distance above which
Chae \cite{chae2024} claimed gravitational anomalies which is $r_{gr~anomaly} \simeq 5$ kau.
\beq
a_{gr~anomaly}  \equiv \frac{G M_\odot}{r_{gr~anomaly}^2} \simeq 2.4~10^{-10}~ {\rm m/s^2}. \label{agranomaly}
\enq
The question of the importance of retardation within the binary system is resolved by
\ern{Fr4} and depends on the relation between the typical size of the system and the 
typical acceleration. For close binaries the critical radius is: 
\beq
 R_{c~max} \simeq \frac{2 c^2}{a_{max}}  \simeq 5~10^{8}~{\rm kau}
\label{Rcmax}
\enq
which is clearly much bigger than a typical size of a close binary system and thus retardation can be neglected for such systems. For wide binaries the importance shifts according to \ern{a12} from internal accelerations to center of mass accelerations. Leading to a much higher critical radius:
 \beq
 R_{c~wide} \simeq \frac{2 c^2}{a_{cm}}  \simeq 4.8~10^{12}~{\rm kau}
\label{Rcwide}
\enq
which is of course much bigger than a wide binary system making the gravitational retardation interaction of all binaries considered negligible. This does mean that gravitational retardation effects are not important for such systems, but what this means is that they are encapsulated in the $\vec F_e$ terms which take into account the gravity of distance stars far away from the binary system. With this in mind we may rewrite \ern{acm} for the center of mass acceleration as:
\beq
\vec a_{cm} = \frac{1}{M}[\vec F_{r1,2} + \vec F_{r2,1} + \vec F_{e1} + \vec F_{e2}]
\simeq \frac{1}{M}[\vec F_{e1} + \vec F_{e2}].
\label{acm2}
\enq 
And also the relative acceleration as:
\beq
\vec a = \vec g_{1,2} - \vec g_{2,1} + \vec g_{e2} - \vec g_{e1}
\simeq \vec g_{N1,2} - \vec g_{N2,1} + \vec g_{e2} - \vec g_{e1}, \qquad
\vec g_{Nj,k} \equiv \frac{\vec F_{Nj,k}}{m_k}.
\label{a2}
\enq
Now:
\beq
g_{N1,2} - \vec g_{N2,1}= \frac{\vec F_{N1,2}}{m_2} - \frac{\vec F_{N2,1}}{m_1}
= \frac{\vec F_{N1,2}}{\mu} = -\frac{G M}{r^2}\hat r, \quad \frac{1}{\mu} \equiv \frac{1}{m_1} + \frac{1}{m_2}, \quad \hat r \equiv \frac{\vec r}{r}.
\label{a3}
\enq
It follows that we can write the relative acceleration as:
\beq
\vec a \simeq  -\frac{G M}{r^2}\hat r + \vec g_{e2} - \vec g_{e1}.
\label{a4}
\enq

\section{Slowly Varying External Gravitational Fields}

Let us estimate the external gravitational acceleration $\vec g_{e1}$ using \ern{r12}, it follows that:
\beq 
\vec g_{e1} = \vec g_{e} (\vec r_1) = \vec g_{e} (\vec r_1) = \vec g_{e} (\vec R_{cm} -\frac{m_2}{M} \vec r).
\label{ge1}
\enq
Now according to \ern{orofmag} $r<<R_{cm}$ so if $\vec g_{e}$ is slowly varying we may approximate:
\beq
\vec g_{e1} \simeq \vec g_{e} (\vec R_{cm}) -\frac{m_2}{M} \vec r \cdot \vec \nabla \vec g_{e} (\vec x)|_{\vec R_{cm}} .
\label{ge12}
\enq
Similarly:
\beq
\vec g_{e2} \simeq \vec g_{e} (\vec R_{cm}) +\frac{m_1}{M} \vec r \cdot \vec \nabla \vec g_{e} (\vec x)|_{\vec R_{cm}} .
\label{ge22}
\enq
Inserting \ern{ge12} and \ern{ge22} into \ern{acm2} we arrive at the simple form:
\beq
\vec a_{cm} \simeq \frac{1}{M}[\vec F_{e1} + \vec F_{e2}] 
 = \frac{1}{M}[m_1 \vec g_{e1} + m_2 \vec g_{e2}] \simeq \vec g_{e} (\vec R_{cm}) 
 \equiv \vec g_{e~cm}.
\label{acm3}
\enq 
Moreover:
\beq
\vec g_{e2} - \vec g_{e1} \simeq \vec r \cdot \vec \nabla \vec g_{e} (\vec x)|_{\vec R_{cm}}.
\label{gedif}
\enq 
Inserting \ern{gedif} into \ern{a4} leads to:
\beq
\vec a \simeq  -\frac{G M}{r^2}\hat r + \vec g_{e2} - \vec g_{e1}
\simeq -\frac{G M}{r^2}\hat r + \vec r \cdot \vec \nabla \vec g_{e} (\vec x)|_{\vec R_{cm}}.
\label{a5}
\enq
We are now at a position to calculate the individual acceleration of each star in the binary
system by plugging \ern{acm3} and \ern{a5} into \ern{a12}
\ber
 \vec a_1  &=&  \vec a_{cm} -\frac{m_2}{M} \vec a
 \simeq \vec g_{e~cm} + \frac{G m_2}{r^2}\hat r -\frac{m_2}{M} \vec r \cdot \vec \nabla \vec g_{e} (\vec x)|_{\vec R_{cm}},
 \nonumber \\
\vec a_2  &=&  \vec a_{cm} +\frac{m_1}{M} \vec a
\simeq \vec g_{e~cm} -\frac{G m_1}{r^2}\hat r + \frac{m_1}{M} \vec r \cdot \vec \nabla \vec g_{e}(\vec x)|_{\vec R_{cm}}.
\label{a12b}
\enr  
Although it is quite clear that:
\beq
g_{e~cm} \gg |\vec r \cdot \vec \nabla \vec g_{e}(\vec x)|_{\vec R_{cm}}|,
\label{gegg}
\enq
let us study a simple model to show this quantitatively. Assume that the mass of the entire galaxy
(excluding the binary system) is located at the origin, that is it is a giant black hole of mass
$M_G$ which is depicted in figure \ref{schembin}, in this case the external acceleration is:
\beq
\vec g_{e} (\vec x) = -\frac{G M_G}{x^3} \vec x,
\label{gex}
\enq
It thus follows that:
\beq
\frac{\partial g_{e~i}}{\partial x_j} = -\frac{G M_G}{x^3} (\delta_{ij}- 3 \frac{x_i x_j}{x^2})
= -\frac{g_e}{x} (\delta_{ij}- 3 \frac{x_i x_j}{x^2}).
\label{gexij}
\enq
in which $\delta_{ij}$ is a Kronecker delta. We can then easily calculate:
\beq
|\vec r \cdot \vec \nabla \vec g_{e}(\vec x)|_{\vec R_{cm}}|
= g_{e~cm}\frac{r}{R_{cm}} |\hat r - 3 (\hat r \cdot \hat R_{cm})\hat R_{cm}|
= g_{e~cm}\epsilon |\hat r - 3 (\hat r \cdot \hat R_{cm})\hat R_{cm}|.
\label{gegg2}
\enq
were we used the definition of $\epsilon$ given in \ern{orofmag}. We notice that:
\beq
|\hat r - 3 (\hat r \cdot \hat R_{cm})\hat R_{cm}|=\sqrt{1+3(\hat r \cdot \hat R_{cm})^2}
\quad \Rightarrow \quad 
1 \le |\hat r - 3 (\hat r \cdot \hat R_{cm})\hat R_{cm}| \le 2.
\label{gegg3}
\enq
Thus:
\beq
|\vec r \cdot \vec \nabla \vec g_{e}(\vec x)|_{\vec R_{cm}}| \le 2 g_{e~cm}\epsilon \ll g_{e~cm},
\label{gegg4}
\enq
as expected. We can thus simplify \ern{a12b} as follows:
\ber
 \vec a_1  &\simeq&  \vec g_{e~cm} + \frac{G m_2}{r^2}\hat r,
 \nonumber \\
\vec a_2  &\simeq& \vec g_{e~cm} -\frac{G m_1}{r^2}\hat r.
\label{a12c}
\enr
This indicates that there is another critical distance (unrelated to retardation per se) below in
which one may neglect all external effects in the system and consider it as "stand alone" in the universe.
\beq
 g_{e~cm} \ll \frac{G m_i}{r^2}, \quad \Rightarrow  \quad r \ll r_{ci} = \sqrt{\frac{G m_i}{g_{e~cm}}}
 \label{rci}
\enq
For a typical star with the sun mass this takes the value:
\beq
r_{c \odot} = \sqrt{\frac{G m_\odot}{g_{e~cm}}} \simeq 5~{\rm kau} \simeq  r_{gr~anomaly}.
\label{rcisun}
\enq
Thus we see again that the scale in which gravitational anomaly is announced is also
the scale in which the galaxy "interferes" in the affairs of specific binary systems which are
wide enough to react to such interference.

\section{Conclusion}

The main results of the current paper are twofold, first we establish a criterion by which slow star systems may suffer retarded gravity effects. This is given by an "uncertainty relation" type formula:
\beq
R~a > 2 c^2
\label{unceretarded}
\enq 
demanding that a star $j$ may affect a star $k$ by retarded gravity only if their distance $R$ times the acceleration of either of them denoted $a$ is larger than twice the speed of light in vacuum square.
This may provide an additional mechanism to retarded gravity in stable galaxies beyond the gas depletion mechanism described in \cite{YaRe3} and the galactic wind mechanism described in \cite{Wagman,Wagman2}. Hence we derived the uncertainty relation of retarded gravity. Second
we have shown that within binary systems, retarded gravity is not important and thus may affect the binary star system only externally (that is by matter which is very far away from the binary stars). We have shown that anomalous gravity is reported only in binary systems in which the external gravitational pull is about the size of the internal gravitational pull. As we do not know the precise coordinates of each binary star but only the projection of its trajectory on the sky this requires careful consideration.


\begin{thebibliography}{999}

\bibitem{Chae2024}
Kyu-Hyun Chae, "Measurements of the Low-Acceleration Gravitational Anomaly from the Normalized Velocity Profile of Gaia Wide Binary Stars and Statistical Testing of Newtonian and Milgromian Theories" arXiv:2402.05720 [astro-ph.GA]

\bibitem[Banik \& Kroupa(2019)]{banik2019} {Banik, I.}, {Kroupa, P.} 2019, \mnras, 487, 1653

\bibitem[Banik et al.(2024)]{banik2024} {Banik, I.}, {Pittordis, C.}, {Sutherland, W.}, {Famaey, B.}, {Ibata, R.}, {Mieske, S.}, {Zhao, H.} 2024, MNRAR, in press (arXiv:2311.03436)

\bibitem[Banik \& Zhao(2018)]{banik2018} {Banik, I.}, {Zhao, H.} 2018, \mnras, 480, 2660

\bibitem[Bekenstein \& Milgrom(1984)]{bekenstein1984} {Bekenstein, J.}, {Milgrom, M.} 1984, \apj, 286, 7

\bibitem[Chae et al.(2022)]{chae2022b} {Chae, K.-H.}, {Lelli, F.}, {Desmond, H.}, {McGaugh, S. S.}, {Schombert, J. M.} 2022, \prd, 106, 103025

\bibitem[Chae(2023a)]{chae2023a} {Chae, K.-H.} 2023a, \apj, 952, 128

\bibitem[Chae(2023b)]{chae2023b} {Chae, K.-H.} 2023b, Python scripts to test gravity with the dynamics of wide binary stars v5, Zenodo

\bibitem[Chae(2024)]{chae2024} {Chae, K.-H.} 2024, \apj, 960, 114

\bibitem[Chae et al.(2021)]{chae2021} {Chae, K.-H.}, {Desmond, H.}, {Lelli, F.}, {McGaugh, S. S.}, {Schombert, J. M.} 2021, ApJ, 921, 104

\bibitem[Chae et al.(2020)]{chae2020b} {Chae, K.-H.}, {Lelli, F.}, {Desmond, H.}, {McGaugh, S. S.}, {Li, P.}, {Schombert, J. M.} 2020, ApJ, 904, 51

\bibitem[Chae \& Milgrom(2022)]{chae2022a} {Chae, K.-H.}, {Milgrom, M.} 2022, \apj, 928, 24

\bibitem[Clarke(2020)]{clarke2020} {Clarke, C. J.} 2020, \mnras, 491, L72

\bibitem[Einstein(1916)]{einstein1916} {Einstein, A.} 1916, Annalen der Physik, 354, 769

\bibitem[El-Badry(2019)]{elbadry2019}  {El-Badry, K.} 2019, \mnras, 482, 5018

\bibitem[El-Badry et al.(2021)]{elbadry2021} {El-Badry, K.}, {Rix, H.-W.}, {Heintz, T. M.} 2021, \mnras, 506, 2269

\bibitem[Famaey \& Binney(2005)]{famaey2005} {Famaey, B.}, {Binney, J.} 2005, \mnras, 363, 603

\bibitem[Foreman-Mackey et al.(2013)]{emcee} Foreman-Mackey, D., Hogg, D. W., Lang, D., Goodman, J. 2013, \pasp, 125, 306

\bibitem[Hernandez(2023)]{hernandez2023} {Hernandez, X.} 2023, \mnras, 525, 1401

\bibitem[Hernandez \& Chae(2023)]{hernandez2024a} {Hernandez, X.}, {Chae, K.-H.} 2023, arXiv:2312.03126

\bibitem[Hernandez et al.(2022)]{hernandez2022} {Hernandez, X.}, {Cookson, S.}, {Cort\'{e}s, R. A. M.} 2022, \mnras, 509, 2304

\bibitem[Hernandez et al.(2019)]{hernandez2019} {Hernandez, X.}, {Cort\'{e}s, R. A. M.}, {Allen, C.}, {Scarpa, R.} 2019, IJMPD, 28, 1950101

\bibitem[Hernandez et al.(2012)]{hernandez2012} {Hernandez, X.}, {Jim\'{e}nez, M. A.}, {Allen, C.} 2012, EPJC, 72, 1884

\bibitem[Hernandez et al.(2024)]{hernandez2024} {Hernandez, X.}, {Verteletskyi, V.}, {Nasser, L.}, {Aguayo-Ortiz, A.} 2024, MNRAS, in press (arXiv:2309.10995)

\bibitem[Hwang et al.(2022)]{hwang2022} {Hwang, H.-C.}, {Ting, Y.-S.}, {Zakamska, N. L.} 2022, \mnras, 512, 3383

\bibitem[McGaugh et al.(2016)]{mcgaugh2016} {McGaugh, S. S.}, {Lelli, F.}, {Schombert, J. M.} 2016, \prl, 117, 201101

\bibitem[Milgrom(1983)]{milgrom1983} {Milgrom, M.} 1983, \apj, 270, 365

\bibitem[Milgrom(2010)]{milgrom2010} {Milgrom, M.} 2010, \mnras, 403, 886

\bibitem[Pecaut \& Mamajek(2013)]{pecaut2013} {Pecaut, M. J.}, {Mamajek, E. E.} 2013, \apjs, 208, 9

\bibitem[Pittordis \& Sutherland(2018)]{pittordis2018} {Pittordis, C.}, {Sutherland, W.} 2018, \mnras, 480, 1778

\bibitem[Pittordis \& Sutherland(2019)]{pittordis2019} {Pittordis, C.}, {Sutherland, W.} 2019, \mnras, 488, 4740

\bibitem[Pittordis \& Sutherland(2023)]{pittordis2023} {Pittordis, C.}, {Sutherland, W.} 2023, OJAp, 6, 4

\bibitem[Shaya \& Olling(2011)]{shaya2011} {Shaya, E. J.}, {Olling, R. P.} 2011, \apjs, 192, 2

\bibitem[Tokovinin(2014)]{tokovinin2014} {Tokovinin, A.} 2014, \aj, 147, 87

\bibitem[Tokovinin(2022)]{tokovinin2022} {Tokovinin, A.} 2022, \apj, 926, 1

\bibitem[Tokovinin \& Kiyaeva(2016)]{tokovinin2016} {Tokovinin, A.}, {Kiyaeva, O.} 2016, \mnras, 456, 2070

\bibitem[Vallenari et al.(2023)]{dr3} {Vallenari, A.}, {Brown, A. G. A.}, {Prusti, T.}, {et al. (Gaia Collaboration)} 2023, \aap, 674, A1

\bibitem[Wall \& Jenkins(2012)]{wall2012} {Wall, J. V.}, {Jenkins, C. R.} 2012, Practical Statistics for Astronomers (Cambridge University Press, 2nd ed.), Section 5.3.1


\bibitem[Babcock(1939)]{Babcock1939} Babcock H. W., 1939, Lick Observatory Bulletin 19, 41

\bibitem[Abbott(2016)]{GwaveObs} Abbott B. P., Abbott R., Abbott T. D. et al., 2016, Phys. Rev.

\bibitem[Castelvecchi(2016)]{Castelvecchi}
Castelvecchi D., Witze W., \emph{Nature News}  \textbf{2016},  doi:10.1038/nature.2016.19361.

\bibitem[de Swart(2017)]{DarkMatterMatter}
de Swart J. G., Bertone G., van Dongen J., Nature Astronomy, 2017, 1, 0059 Macmillan Publishers Limited https://doi.org/10.1038/s41550-017-0059

\bibitem[Eddington(1923)]{Edd}
Eddington A. S. , "The mathematical theory of relativity" Cambridge University Press (1923)

\bibitem[Einstein(1916)]{Einstein2}
Einstein A., \emph{Sitzungsberichte der Königlich Preussischen Akademie der Wissenschaften Berlin}; Part 1;  \textbf{1916}; pp. 688–696. The Prusssian Academy of Sciences, Berlin, Germany.

\bibitem[Feynman(2011)] {Feynman}
Feynman R. P.,  Leighton R. B., Sands M. L., Feynman Lectures on Physics, Basic Books; revised 50th anniversary edition (2011).

\bibitem [Jackson(1999)]{Jackson}
Jackson J. D., Classical Electrodynamics, Third Edition. Wiley: New York, (1999).

\bibitem[Landau(1975)]{LandLiftFields}
Landau L. D., 1975, The classical theory of fields, 4th edn. (Pergamon)

\bibitem[Mannheim(1993)]{Mann93}
Mannheim P. D., 1993, The Astrophysical Journal, 419, 150

\bibitem[Mannheim(1996)]{ManExt}
Mannheim P. D., 1996, Foundations of Physics, 26, 1683

\bibitem [McGaugh(2017)]{McGaugh2017}
 McGaugh S., McGaugh's Data Pages. N.p., n.d. 2017. Available online: http://astroweb.case.edu/ssm/data/ (accessed on 22 January 2017).

\bibitem[Milgrom(1983)]{Milg1983}
Milgrom M., 1983, The Astrophysical Journal, 270, 371

\bibitem[Moffat(2006)]{Moffat}
Moffat J. W., (2006).  Journal of Cosmology and Astroparticle Physics. 2006 (3): 4. arXiv:gr-qc/0506021. doi:10.1088/1475-7516/2006/03/004.

\bibitem[Misner(1973)]{MTW}
 Misner C. W.,  Thorne K.S., Wheeler J.A., "Gravitation" W.H. Freeman \& Company (1973)

 \bibitem[Narlikar(1993)]{Narlikar}
Narlikar, J. V. (1993). Introduction to Cosmology, Second Edition. Cambridge University Press.

\bibitem[Navarro(1996)]{Navarro}
Navarro J. F., Frenk C. S., White S. D. M., (May 10, 1996). The Astrophysical Journal. 462: 563-575. arXiv:astro-ph/9508025. doi:10.1086/177173.

\bibitem[Taylor(1993)]{Taylor}
Nobel Prize, A. \emph{Press Release The Royal Swedish Academy of Sciences}; \textbf{1993}.The Royal Swedish Academy of Sciences, Stockholm, Sweden.

\bibitem[Rubin(1970)]{rubin1}
Rubin V.C., Ford W.K. Jr., {\it Astrophys. J.}, vol. 159, 379, 1970.

\bibitem[Rubin(1980)]{rubin2}
Rubin V.C.,  Ford W.K. Jr., Thonnard N., {\it Astrophysical Journal}, vol. 238, 471, 1980.

\bibitem[Sancisi(2003)]{Sancisi}
Sancisi R., proceedings of IAU Symposium 220, "Dark Matter in Galaxies", eds. S. Ryder, D.J. Pisano, M. Walker and K. Freeman, Publ. Astron. Soc. Pac arXiv:astro-ph/0311348

 \bibitem [Sanders(2002)]{Sanders2002}
Sanders R., McGaugh S., Annu. Rev. Astron. Astrophys. 2002, 40, 217.

\bibitem[Schwinger(1998)]{Schwinger}
Schwinger J., Lester L., DeRaad Jr K. W., 1998, Classical Electrodynamics,
Advanced Book Program (Reading, Massacusetts: Perseus Books).

\bibitem[Tully(1977)]{TF}
Tully R. B., Fisher J. R., (1977).  Astronomy and Astrophysics. 54 (3): 661–673.

\bibitem[van Dokkum(2018)]{Dokkum}
van Dokkum P.,  Danieli S.,  Cohen Y.,  Merritt A.,  Romanowsky A.J.,  Abraham R.,  Brodie J., Conroy C.,   Lokhorst D.,  Mowla L., ~et~al. \emph{Nature} \textbf{2018}, \emph{555}, 629–632,
doi:10.1038/nature25767.

\bibitem[Volders(1959)]{volders}
  Volders L.M.J.S., {\it Bull. astr. Inst. Netherl.}, vol. 14, 323, 1959.
  Rubin V.C.,  Ford Jr. W.K.,  Thonnard N., and Roberts M.S., {\it Astrophys. J.}, vol. 81, 687 and 719, 1976.

\bibitem[Wagman(2019)]{Wagman}
Wagman M.,  Retardation Theory in Galaxies. Ph.D. Thesis, Senate of Ariel University, Samria, Israel, 23~September 2019.

\bibitem[Wagman(2023)]{Wagman2}
Wagman M.,  Horwitz L. P., Yahalom A., 2023 J. Phys.: Conf. Ser. 2482 012005.  Proceedings of the 13th Biennial Conference on Classical and Quantum Relativistic Dynamics of Particles and Fields (IARD 2022), 05/06/2022 - 09/06/2022 Prague, Czechia. DOI 10.1088/1742-6596/2482/1/012005.

\bibitem[Weinberg(1972)]{Weinberg}
 Weinberg S., "Gravitation and Cosmology: Principles and Applications of the General Theory of Relativity" John Wiley \& Sons, Inc. (1972)

 \bibitem[Yahalom(2008)]{Yahalom}
Yahalom A., Foundations of Physics, Volume 38, Number 6, Pages 489-497 (June 2008).

\bibitem[Yahalom(2009)]{Yahalomb}
Yahalom A., (2009). International Journal of Modern Physics D, Vol. 18, Issue: 14, pp. 2155-2158.

 \bibitem[Yahalom(2018)]{ge}
Yahalom A.,  Retardation Effects in Electromagnetism and Gravitation. In Proceedings of the Material Technologies and Modeling the Tenth International Conference, Ariel University, Ariel, Israel,  20–24 August~2018. (arXiv:1507.02897v2)

\bibitem[Yahalom(2019a)]{YaRe2}
Yahalom A., Dark Matter: Reality or a Relativistic Illusion? In Proceedings of Eighteenth Israeli-Russian Bi-National Workshop 2019, Ein Bokek, Israel, 17--22 February 2019.

 \bibitem[Yahalom(2019b)]{YaRe1}
Yahalom A., J. Phys.: Conf. Ser. 1239 (2019) 012006.

\bibitem[Yahalom(2020)]{YaRe3}
 Yahalom A., Symmetry 2020, 12(10), 1693; https://doi.org/10.3390/sym12101693

\bibitem[Yahalom(2021a)]{YaRe5}
Yahalom A., Proceedings of IARD 2020. 2021 J. Phys.: Conf. Ser. 1956 012002

\bibitem[Yahalom(2021b)]{YaRe4}
Yahalom A., Universe. 2021; 7(7):207. https://doi.org/10.3390/universe7070207. https://arxiv.org/abs/2108.08246

\bibitem[Yahalom(2021c)]{lensing}
Yahalom A., Symmetry 2021, 13, 1062.  https://doi.org/10.3390/sym13061062. https://arxiv.org/abs/2108.04683.

\bibitem[Yahalom(2021d)]{YaRe6}
Yahalom A., International Journal of Modern Physics D, (2021), Volume No. 30, Issue No. 14, Article No. 2142008 (8 pages). \copyright World Scientific Publishing Company.https://doi.org/10.1142/S0218271822420184

\bibitem[Yahalom(2022a)]{YaRe9}
Yahalom A.,  Proceedings of the International Conference: COSMOLOGY ON SMALL SCALES 2022 Dark Energy and the Local Hubble Expansion Problem, Prague, September 21-24, 2022. Edited by Michal Krizek and Yuri V. Dumin, Institute of Mathematics, Czech Academy of Sciences.

\bibitem[Yahalom(2022b)]{lensing2}
Yahalom A., IJMPD Vol. 31, No. 14, 2242018 (10 pages), received 23 May 2022, Accepted 31 August 2022, published online 30 September 2022.

\bibitem[Yahalom(2023a)]{YaRe9b}
Yahalom A., Symmetry 2023, 15, 39. https://doi.org/10.3390/sym15010039

 \bibitem[Yahalom(2023b)]{YaRe10}
Yahalom A.,  International Journal of Modern Physics D, 2342013, Received 12 May 2023, Accepted 3 July 2023, Published: 28 July 2023. https://doi.org/10.1142/S0218271823420130, https://www.worldscientific.com/doi/abs/10.1142/S0218271823420130

\bibitem[Yahalom(2023c)]{YaRe11}
Yahalom A., accepted by Bulgarian Journal of Physics vol. 50 (2023) 1–16.

\bibitem[Zwicky(1937)]{zwicky}
 Zwicky F., In: Proc. Natl. Acad. Sci. U S A., May 1937, vol. 23(5),  pp. 251–256.

 \bibitem {Tuval}
Tuval, M.; Yahalom, A. Newton's Third Law in the Framework of Special Relativity. \emph{Eur. Phys. J. Plus} \textbf{2014}, \emph{129}, 240, doi:10.1140/epjp/i2014-14240-x.
\bibitem {YahalomT}
Tuval, M.; Yahalom, A.   Momentum Conservation in a Relativistic Engine. \emph{Eur. Phys. J. Plus} \textbf{2016}, \emph{131}, 374,
doi:10.1140/epjp/i2016-16374-1.
\bibitem {Yahalom3}
Yahalom, A.  Retardation in Special Relativity and the Design of a Relativistic Motor. \emph{Acta Phys. Pol. A}  \textbf{2017}, \emph{131}, 1285--1288.
\bibitem {Yahalom4}
Shailendra Rajput, Asher Yahalom \& Hong Qin "Lorentz Symmetry Group, Retardation and Energy Transformations in a Relativistic Engine" Symmetry 2021, 13, 420. https://doi.org/10.3390/sym13030420.
\bibitem {Yahalom5}
Rajput, Shailendra, and Asher Yahalom. 2021. "Newton's Third Law in the Framework of Special Relativity for Charged Bodies" Symmetry 13, no. 7: 1250. https://doi.org/10.3390/sym13071250
\bibitem {Yahalom6}
Yahalom, Asher. 2022. "Newton's Third Law in the Framework of Special Relativity for Charged Bodies Part 2: Preliminary Analysis of a Nano Relativistic Motor" Symmetry 14, no. 1: 94. https://doi.org/10.3390/sym14010094.

\end{thebibliography}
\end{document}